\documentstyle[preprint,aps,epsf]{revtex}                  
\tightenlines
\begin{document}
\def\non{\nonumber}
\def\be{\begin{eqnarray}}
\def\en{\end{eqnarray}}
\def\la{\langle}
\def\ra{\rangle}
\def\hep{\hat{\varepsilon}}
\def\pr{Phys. Rev.~}
\def\prl{Phys. Rev. Lett.~}
\def\pl{Phys. Lett.~}
\def\np{Nucl. Phys.~}
\def\zp{Z. Phys.~}
\def\bi{\bibitem}
\pagestyle{empty}                                      
\draft
\vfill
\title{Consistent treatment for valence and nonvalence configurations \\
in semileptonic weak deacys}

\author{Chien-Wen Hwang} %
\address{\rm Department of Physics, National Tsing Hua University, Hsinchu 300, Taiwan \rm}
%
%
\vfill
\maketitle
\begin{abstract}
We discuss the semileptonic weak decays of $P\to P$ ($P$ denotes a
pseudoscalar meson). In these timelike processes, the problem of
the nonvalence contribution is solved systematically as well as
the valence one. These contributions are related to the
light-front quark model (LFQM), and the numerical results show the
nonvalence contribution of the light-to-light transition is larger
than of the heavy-to-light one. In addition, the relevant CKM
matrix elements are calculated. They are consistent with the data
of Particle Data Group.
\end{abstract}
%
\pacs{PACS numbers: 13.20.-v, 12.39.Ki}
%
\pagestyle{plain}
The study of exclusive semileptonic decays has
attracted much interest for a long time. Heavy-to-heavy
semileptonic decays, such as $B \to Dl\nu$, provide an ideal
testing ground for heavy-quark symmetry and heavy-quark effective
theory (for a review, see \cite{Neubert}). On the other hand,
heavy-to-light and light-to-light weak decays are much more
complicated theoretically since there exists no guiding symmetry
principle. Nevertheless, it is essential to understand the
reaction mechanisms of these decay modes, because they are the
main sources of information on the CKM mixing matrix between heavy
and light quarks.

Hadronic matrix elements of weak $P \to P$ transition is described
by two form factors. Phenomenologically, the hadronic form factors
can be evaluated in various models, including the popular quark
model. However, since usual quark-model wave functions best
resemble meson states in the rest frame, or where the meson
velocities are small, the form factors calculated in the
non-relativistic quark model are therefore trustworthy only when
the recoil momentum of the daughter meson relative to the parent
meson is small. As the recoil momentum increases (corresponding to
decreasing $q^2$), we have to consider relativistic effects
seriously.

It is well known that the LFQM \cite{Ter,Chung} is a relativistic
quark model in which a consistent and fully relativistic treatment
of quark spins and the center-of-mass motion can be carried out.
This model has many advantages. For example, the light-front (LF)
wave function is manifestly Lorentz invariant as it is expressed
in terms of the momentum fraction variables (in ``+" components)
in analogy with the parton distributions in the infinite momentum
frame. Moreover, hadron spin can also be correctly constructed
using the so-called Melosh rotation. The kinematic subgroup of the
LF formalism has the maximum number of interaction-free
generators, including the boost operator which describes the
center-of-mass motion of the bound state (for a review of LF
dynamics, see \cite{Zhang}). The LFQM has been applied in the past
to study the heavy-to-heavy and heavy-to-light weak decay form
factors \cite{Jaus,Don94}. However, the weak form factors were
calculated only for $q^2\leq 0$ at the beginning, whereas physical
decays occur in the time-like region $0\leq q^2\leq (M_i-M_f)^2$,
with $M_{i,f}$ being the initial and final meson masses. Hence
extra assumptions are needed to extrapolate the form factors to
cover the entire range of momentum transfer \cite{Jaus96,Mel}.
Lately, the weak form factors for $P\to P$ transition were
calculated in \cite{Sima,Cheung2} for the first time for the
entire range of $q^2$, so additional extrapolation assumptions are
no longer required. This is based on the observation \cite{Dubin}
that in the frame where the momentum transfer is purely
longitudinal i.e. $q_\perp=0$, $q^2=q^+q^-$ covers the entire
range of momentum transfer. The price is that, besides the
conventional valence-quark contribution, we must also consider the
nonvalence configuration (or the so-called Z graph, see FIG.~1
(b)). The nonvalence contribution vanishes if $q^+=0$, but is
supposed to be important for heavy-to-light transition near zero
recoil \cite{Jaus,Jaus96,Dubin,Saw}. Some methods for treating
this nonvalence configuration exist: the authors of Ref.
\cite{Cheung2} considered the effective higher Fock state and
calculated the effect in chiral perturbation theory. Ref.
\cite{CJ} follow a Schwinger-Dyson approach and related the
nonvalence contribution to an ordinary LF wave function.

In this letter, we present a new way of handling the nonvalence
contribution of $P \to P$ transition. For comparsion, it will be
instructive to analyze the known valence contribution in parallel.
The main advantage of this way is that relativistic effects of the
quark motion and spin are treated consistently in both valence and
nonvalence configurations. We assume both normalization conditions
of meson and quark states and a single interaction Hamiltonian to
obtain both the Melosh transformations of valence and nonvalence
contributions. Combining these two contributions, we calculate
completely the form factors of the semileptonic decay and the
relevant CKM matrix elements.

We are interested in the matrix element which defines the weak
form factors by
\be
\la P'|\bar {Q'} \gamma^\mu Q|P
\ra=f_+(q^2)(P+P')^\mu+f_-(q^2)~q^\mu, \label{PPform}
\en
where $q=P-P'$ is the momentum transfer. Assuming a vertex
function $\Lambda_P$ \cite{Jaus,Don94} which is related to $Q\bar
q$ bound state of $P$ meson, the quark-meson diagram depicted in
FIG.~1 (a) yields
\be
\la P'|\bar {Q'} \gamma^\mu Q|P \ra=-\int {d^4 p_1 \over{(2
\pi)^4}} \Lambda_P \Lambda_{P'}{\rm Tr}\Bigg[\gamma_5
        {i(\not{\! p_3}+m_3)\over{p_3^2-m^2_3+i\epsilon}} \gamma_5
       {i(\not{\!
        p_2}+m_2)\over{p_2^2-m^2_2+i\epsilon}}\gamma^\mu
        {i(\not{\! p_1}+m_1)\over{p_1^2-m^2_1+i\epsilon}}\Bigg],
        \label{amp}
\en
where $p_2=p_1-q$ and $p_3=p_1-P$. We consider the poles in
denominators in terms of the LF corrdinates $(p^-,p^+,p_\perp)$
and perform the integration over the LF ``energy" $p_1^-$ in Eq.
(\ref{amp}). The result is then
\be
\la P'|\bar {Q'} \gamma^\mu Q|P \ra &=& \int^q_0 [d^3
p_1]
           {\Lambda_P \over{p^-_3-p^-_{3{\rm on }}}}(I^\mu|_{p^-_{1{\rm on}}})
           {\Lambda_{P'} \over{p^-_3-p^-_{3{\rm on }}+p^-_2-p^-_{2{\rm on}}}}
           \non \\
&+&\int^P_q  [d^3 p_1]
          {\Lambda_P \over{p^-_1-p^-_{1{\rm on}}}}(I^\mu|_{p^-_{3{\rm on}}})
            {\Lambda_{P'} \over{p^-_2-p^-_{2{\rm on}}}},
            \label{pole}
\en
where $i=1,2,3$, 
\be
&&[d^3 p_1]=dp^+_1 d^2p_{1\perp}/(16\pi^3 \prod_i p^+_i), \non \\
&&I^\mu={\rm Tr}[\gamma_5(\not{\!
p_3}+m_3)\gamma_5(\not{\! p_2}+m_2)\gamma^\mu(\not{\! p_1}+m_1)], \non \\
&&p^-_{i{\rm on}}=m^2_i+p^2_{i\perp}/p^+_i, p^-_{1(3)}=P^-_{\rm
on}-p^-_{3(1){\rm on}}, 
\en
and $p^-_2$ equals respectively
$p^-_{3{\rm on}}-P'^-_{\rm on}$ and $P'^-_{\rm on}-p^-_{3{\rm
on}}$ in the first and second term of Eq. (\ref{pole}). It is
worthwhile to mention every vertex function and its denominator
corresponds exactly to the relevant meson bound state. This is
clearer if we define $S_j \equiv p^-_j-p^-_{j{\rm on}}$ and
rewrite Eq. (\ref{pole}) in a more symmetrical form:
\be
\la P'|\bar {Q'} \gamma^\mu Q|P \ra &=& \int^q_0 [d^3
p_1]
          \Bigg[{\Lambda_P \over{S_P+S_1+S_3}}I^\mu
                {\Lambda_{P'} \over{S_{P'}+S_2+S_3}}\Bigg]
                \Bigg |_{S_{P,P',1}=0} \non \\
&+&\int^P_q [d^3 p_1]
          \Bigg[{\Lambda_P \over{S_P+S_1+S_3}}I^\mu
                {\Lambda_{P'} \over{S_{P'}+S_2+S_3}}\Bigg]
                \Bigg |_{S_{P,P',3}=0}.  \label{SSS}
\en

In general, the integrals in Eq. (\ref{SSS}) are divergent if we
treat the vertices as pointlike. Internal structures for these
vertices are therefore necessary. In the LFQM, the internal
structure \cite{Cheung2,cheng,hwcw2} consists of $\phi$ which
describes the momentum distribution of the constituents in the
bound state, and $R^{S,S_z}_{\lambda_1,\lambda_2}$ which creates a
state of definite spin ($S,S_z$) out of LF helicity
($\lambda_1,\lambda_2$) eigenstates and is related to the Melosh
transformation \cite{Melosh}. Here we adopt a convenient approach
relating these two parts. The interaction Hamiltonian is assumed
to be $H_I=i\int d^3 x \bar {\Psi}\gamma_5 \Psi \Phi $ where
$\Psi$ is quark field and $\Phi$ is meson field containing $\phi$
and $R^{S,S_z}_{\lambda_1,\lambda_2}$. On the one hand, if we
normalize the meson state depicted in FIG.2 (a) as \cite{Cheung2}
\be
\la M (P',S',S'_z)|H_I~H_I|M(P,S,S_z)\ra =2(2\pi)^3P^+\delta^3(P'-P)\delta_{SS'}\delta_{S_zS'_z},
\en
and the valence wave function $\phi^{\rm v}$ as
\be
\int {d^3p_1 \over{2(2\pi)^3}} {1\over{P^+}} |\phi^{\rm v}|^2=1,
\en
where $p_1$ and $p_2$ are the on-mass-shell momenta; the valence
configuration of $R^{S,S_z}_{\lambda_1,\lambda_2}$ is
\be
R^{\rm v}_{1,2}={\sqrt{P^+p^+_1 p^+_2}
                  \over{2\sqrt{p_{1{\rm on}}\cdot p_{2{\rm on}}+m_1 m_2}}}.
\en
On the other hand, if we normalize the quark state
depicted in FIG.2 (b) as
\be
\la Q (p'_3,s')|H_I~H_I|Q(p_3,s)\ra
=2(2\pi)^3\delta^3(p'_3-p_3)\delta_{s's},
\en
and the nonvalence wave function $\phi^{\rm n}$ as
\be
\int {d^3p_2 \over{2(2\pi)^3}} {1\over{p_3^+}} |\phi^{\rm n}|^2=1;
\en
the nonvalence configuration of $R^{S,S_z}_{\lambda_1,\lambda_2}$
is
\be
R^{\rm n}_{2,3}={\sqrt{P'^+p^+_2 p^+_3}
                  \over{2\sqrt{p_{2{\rm on}}\cdot p_{3{\rm on}}-m_2 m_3}}}.
\en
After taking the ``good " component $\mu=+$, the wave function and
the Melosh transformation of the meson are related to the bound
state vertex function $\Lambda_P$ by
\be
&&{\Lambda_P \over{S_P+S_1+S_3}}\Bigg|_{S_{P,P',1}=0}
\longrightarrow R^{\rm v}_{1,3}~\phi^{\rm v}_P, \non \\
&&{\Lambda_{P'}\over{S_{P'}+S_2+S_3}} \Bigg|_{S_{P,P',1}=0}
\longrightarrow R^{\rm n}_{2,3}~\phi^{\rm n}_{P'}.
\en
In the trace of $I^+$, $p_1$, $p_2$, and $p_3$ must be on the mass
shell for self-consistency. Hence the matrix element in LFQM is
\be
\la P'|\bar {Q'} \gamma^+ Q|P \ra &=&\int^q_0 [d^3p_1]
           \left[R^{\rm v}_{1,3}~\phi^{\rm v}_P~(I^+)~
           R^{\rm n}_{2,3}~\phi^{\rm n}_{P'}\right]|_{S_{1,2,3}=0} \non \\
&+&\int^P_q [d^3p_1]
            \left[R^{\rm v}_{1,3}~\phi^{\rm v}_P~(I^+)~R^{\rm
            v}_{2,3}~\phi^{\rm v}_{P'}\right] |_{S_{1,2,3}=0}.
            \label{LFEq}
\en
We use the definitions of the LF momentum variables
$(x,x',k_\perp,k'_\perp)$ \cite{cheng} and take a Lorentz frame
where $P_\perp=P'_\perp=0$ amounts to having $q_\perp=0$ and
$k'_\perp=k_\perp$. So from Eq. (\ref{LFEq}) we obtain
\be
H(r)= \int {d^2k_\perp\over 2(2\pi)^3}
&&\Bigg\{\int^{r}_0dx~ \phi^{\rm v}_P(x,k_\perp) \phi^{\rm
v}_{P'}(x',k_\perp)\,{{\cal A}{\cal A}'+k^2_\perp\over\sqrt{{\cal
A}^2 +k_\perp^2}\sqrt{{\cal A}'^2+k_\perp^2}}\non \\
&&+\int^{1}_rdx~ \phi^{\rm v}_P(x,k_\perp) \phi^{\rm
n}_{P'}(x',k_\perp) \,{{\cal A}{\cal A}'+k^2_\perp\over\sqrt{{\cal
A}^2 +k_\perp^2}\sqrt{{\cal A}'^2+k_\perp^2}} \Bigg\},
\label{Hi}
\en
where 
\be
\la P'|\bar {Q'} \gamma^+ Q |P \ra=2P^+H(r),
\en
${\cal A}=m_1x+m_3(1-x)$, and ${\cal A}'=m_2x'+m_3(1-x')$. $x~(x')$ is
the momentum fraction carried by the spectator antiquark in the
initial (final) state in the first term of (\ref{Hi}). However,
$x' \geq 1$ the second term of (\ref{Hi}), which shows that the
momentum $p^+_3$ of the spectator quark is larger than the $P'^+$
of the final meson.

As explained above, we shall work in the frame where $q_\perp=0$
so that $q^2\geq 0$. Defining $r\equiv P'^+/P^+$ gives $
q^2=(1-r)(M_P^2-M_{P'}^2/r)$. Consequently, for a given $q^2$, the
two solutions for $r$ are given by
\be
r_{\pm}={1\over {2M^2_P}}\big[M_P^2+M_{P'}^2-q^2\pm 2M_p{\cal
Q}(q^2)~\big], \label{y12}
\en
where ${\cal Q}(q^2)=\sqrt{(M^2_P+M^2_{P'}-q^2)^2-4 M^2_P
M^2_{P'}}/2M_P$. The $\pm$ signs in (\ref{y12}) correspond to the
daughter meson recoiling in the $\pm z$-direction relative to the
parent meson. The form factors $f_\pm(q^2)$ of course should be
independent of the reference frame chosen for the moving direction
of the daughter meson. For a given $q^2$, it follows from
(\ref{PPform}) that
\be
f_{\pm}(q^2) &=&\pm {(1 \mp r_-)H(r_+)-(1 \mp r_+)H(r_-)\over
r_+-r_-}\,. \label{fpm}
\en
It is easily seen that $f_\pm(q^2)$ are independent of the choice
of reference frames, as it should be. The scalar form factor
$f_0(q^2)$ is related to $f_\pm(q^2)$ by
\be
f_0(q^2)=f_+(q^2)+{q^2\over{M_P^2-M_{P'}^2}}f_-(q^2). \label{f0}
\en
The differential decay rate for $P\to P$ is given by \cite{CJ}
\be
{d\Gamma\over{dq^2}}&=&{G^2_F\over{24\pi^3}}|V_{q_1 \bar{q}_2}|^2
{\cal Q}(q^2)(1-2\hat {s})^2 \non \\
&\times&\Bigg\{[{\cal
Q}(q^2)]^2(1+\hat {s})|f_+(q^2)|^2
+M^2_P\Bigg(1-{M_{P'}^2\over{M_P^2}}\Bigg)^2{3\over{4}}\hat
{s}|f_0(q^2)|^2\Bigg\}, \label{VV}
\en
where $G_F$ is the Fermi constant, $\hat {s}=m^2_l/2q^2$, $m_l$ is
the mass of lepton $l$, and $V_{q_1 \bar{q}_2}$ is the CKM matrix
element.

In principle, the momentum distribution amplitude $\phi(x,k_\bot)$
can be obtained by solving the LF QCD bound state
equation\cite{Zhang}. However, before such first-principle
solutions are available, we shall have to use phenomenological
amplitudes. The simplest conjecture is related to the Melosh
transformation effect; for example, $\phi=N {\rm exp}[-({\cal A}^2
+k_\perp^2)/(2\omega^2)]$, where $N$ is normalization constant and
$\omega$ is a scale parameter. However, the contributions of the
end-point regions ($x \to 0,1$) for this wave function are
nonvanishing. Here we make a slight modification to:
\be
\phi(x,k_\perp)=N [x(1-x)]^{1/n}\Bigg[{\omega^2\over{({\cal A}^2
+k_\perp^2)+\omega^2}}\Bigg]^n, \label{power}
\en
where $n$ is an integer. When $n$ is large ($\sim 20$), the form
of this power-law wave function is almost the same as the previous
exponential one except at the end-points. In addition, we do not
treat $n$ as a new parameter because the differences between wave
functions for different large $n$'s are negligible. Thus the three
parameters are $m_1$, $m_2$, and $\omega$ in Eq. (\ref{power}). We
can use Eqs. (\ref{fpm}), (\ref{f0}), (\ref{Hi}), and
(\ref{power}) to calculate the form factors of the processes
$K^0\to \pi^\pm l^\mp \nu_l (K^0_{e3})$ and $D^0(B^0) \to \pi^-
l^+ \nu_l$ which correspond to the light-to-light and
heavy-to-light decay modes, respectively. On the one hand, the
parameters appearing in the wave functions $\phi^{\rm v}_{K,\pi}$
are fixed by assuming the quark masses $m_u=m_d$ and fitting to
the experimental values of the decay constants $f_{K,\pi}$
\cite{PDG00} and the charged radii $\la r^2 \ra _{K^+,\pi^+}$
\cite{expKpi,hwcw3}. On the other hand, we determine the parameter
$\omega$ in $\phi^{\rm n}_\pi$ by fitting the data in Ref.
\cite{kpiexp} and treat it as universal among the other decay
modes. As for the $D$ and $B$ mesons, the parameters are
determined by assuming the quark masses $m_c=1.3$ GeV, $m_b=4.5$
GeV and fitting to the lattice QCD values of the decay constant
$f_{D,B}$ \cite{lattice}. These parameters are as listed below (in
units of GeV):
\be
&&m_{u,d}=0.2,~m_s=0.32,~m_c=1.3,~m_b=4.5,~\omega^{\rm n}_\pi=0.3,
\non \\ &&\omega^{\rm v}_\pi=2.34,~\omega^{\rm v}_K=2.66,
~\omega^{\rm v}_D=3.19,~\omega^{\rm v}_B=4.71.
\en
The numerical results of the form factor $f_+$ for
various decay modes are ploted in FIG. 3, 4, 5. From these
figures, we easily find, for the same final meson, that the
nonvalence contributions are smaller when the inital mesons are
heavier. In addition, the nonvalence contribution is important for
heavy-to-light transition near zero recoil ($q^2\sim q^2_{max}$).
This result is consistent with the prediction in
\cite{Jaus,Jaus96,Dubin,Saw}.

Finally, we can use the Eqs. (\ref{fpm}), (\ref{f0}), (\ref{VV})
and the experimental data of the relevant decay rates \cite{PDG00}
to calculate the three CKM matrix elements $V_{us}$, $V_{cd}$, and
$V_{ub}$. These values from this work and Ref. \cite{PDG00} are
listed in Table I. The error bars in this work come from the
uncertainities of the decay widths. We find these values are
consistent with \cite{PDG00}.

\begin{center}
{\small Table 1: The values of some CKM matrix elements in this work and
Ref. [17].}
\end{center}
\begin{center}
\begin{tabular}{|c|c|c|c|}\hline
           & $V_{us}$ & $V_{cd}$ & $V_{ub}$ \\ \hline
 This work  & $0.2179 \pm 0.0016$ & $0.247 \pm 0.021$ & $0.0037 \pm 0.0007$ \\ \hline
 P.D.G.[17] & $0.2196 \pm 0.0023$ & $0.224 \pm 0.016$ & $0.0036
 \pm 0.0012$ \\ \hline
\end{tabular}
\end{center}

In conclusion, a new treatment for the nonvalence configuration
have been shown. We emphasize that the vertex functions correspond
to LF valence and nonvalence wave functions exactly. The
relativistic effects of the quark motion and spin were also
treated consistently in both valence and nonvalence
configurations. Therefore, we are able to calculate the form
factors of the semileptonic decay completely. The numerical
results showed the nonvalence contribution of the heavy-to-light
transition is smaller than that of the light-to-light one. In
addition, the CKM matrix elements evaluated from these form
factors were consistent with the data in Particle Data Group.
\acknowledgments
I would like to thank C.P. Soo for helpful comments and Frederic Blanc for useful data. This work was
supported in part by the National Science Council of ROC under
Contract No. NSC90-2112-M-007-040.


\newpage
\parindent=0 cm
\centerline{\bf FIGURE CAPTIONS}
\vskip 0.5 true cm

{\bf Fig. 1 } (a) The Feynman triangle
diagram. (b) corresponds to the LF nonvalence configuration and
diagram (c) to the valence one. Filled and empty circles incidate
vertex functions and LF wave functions respectively.
\vskip 0.25 true cm

{\bf Fig. 2 } The Feynman diagrams of
the self-energy of (a) meson and (b) quark.

{\bf Fig. 3 } The normalized form factor
$F_+$ for $K^0_{e3}$ decay compared with the experimental data
[21]. The definition of $F_+$ is $f_+(q^2)/f_+(0)$.

{\bf Fig. 4 } The form factor $f_+$ for
$D^0 \to \pi^- l^+ \nu_l$ compared with the lattice QCD data
[22].

{\bf Fig. 5 } The form factor $f_+$ for
$B^0 \to \pi^- l^+ \nu_l$ compared with the lattice QCD data
[23].

\newpage

\begin{figure}[h]
\includegraphics{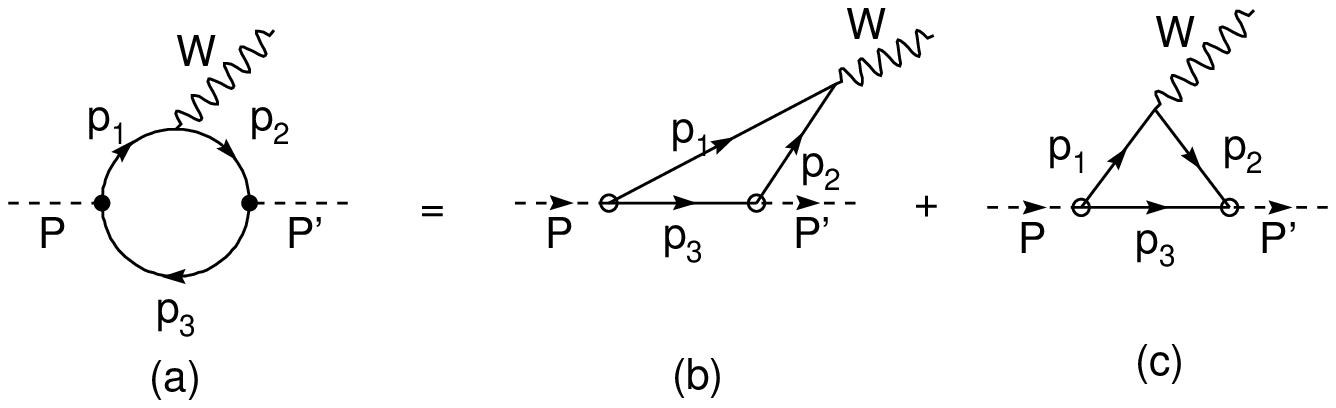}
\caption{}
\vskip 13.cm
\end{figure}
\newpage

\begin{figure}[h]
\includegraphics{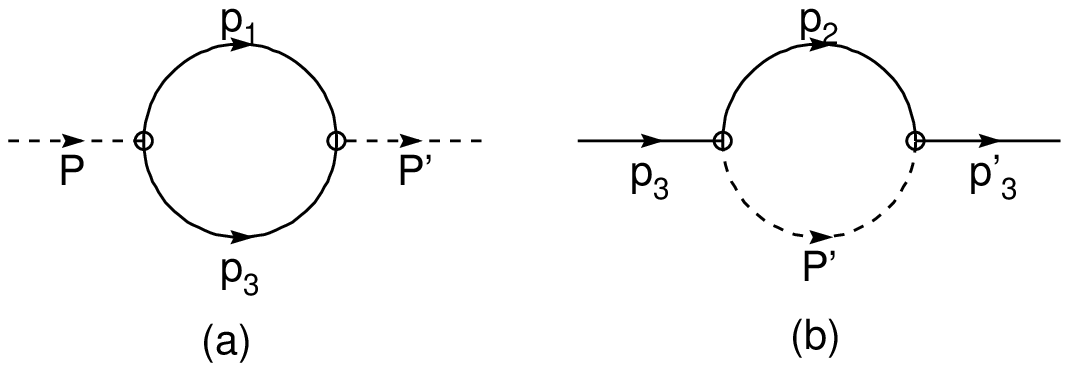}
\caption{}
\vskip 13.cm
\end{figure}
\newpage

\begin{figure}[h]
\includegraphics{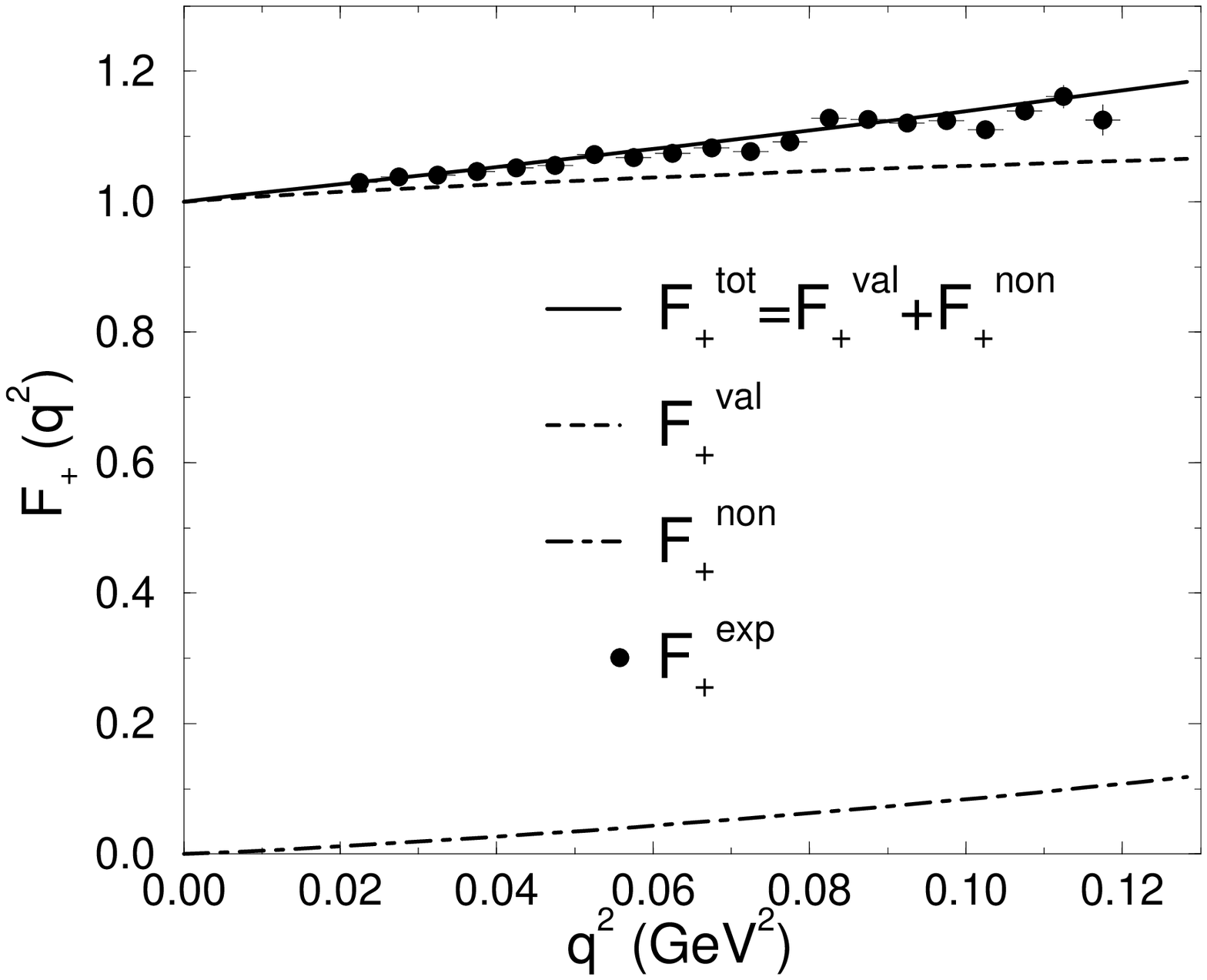}
\vskip 23.cm
\caption{}
\end{figure}
\newpage

\begin{figure}[h]
\includegraphics{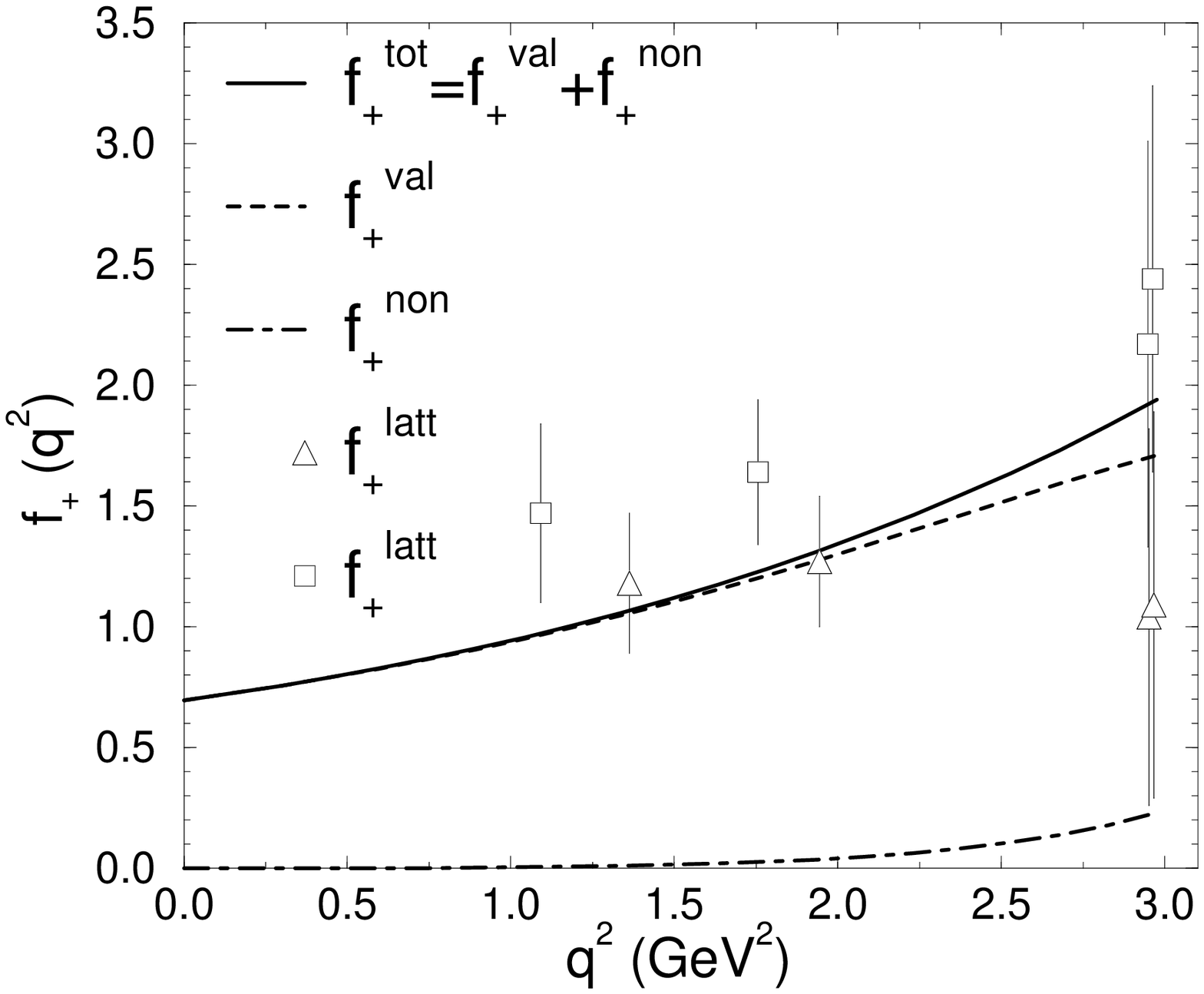}
\vskip 23.cm
\caption{}
\end{figure}
\newpage

\begin{figure}[h]
\includegraphics{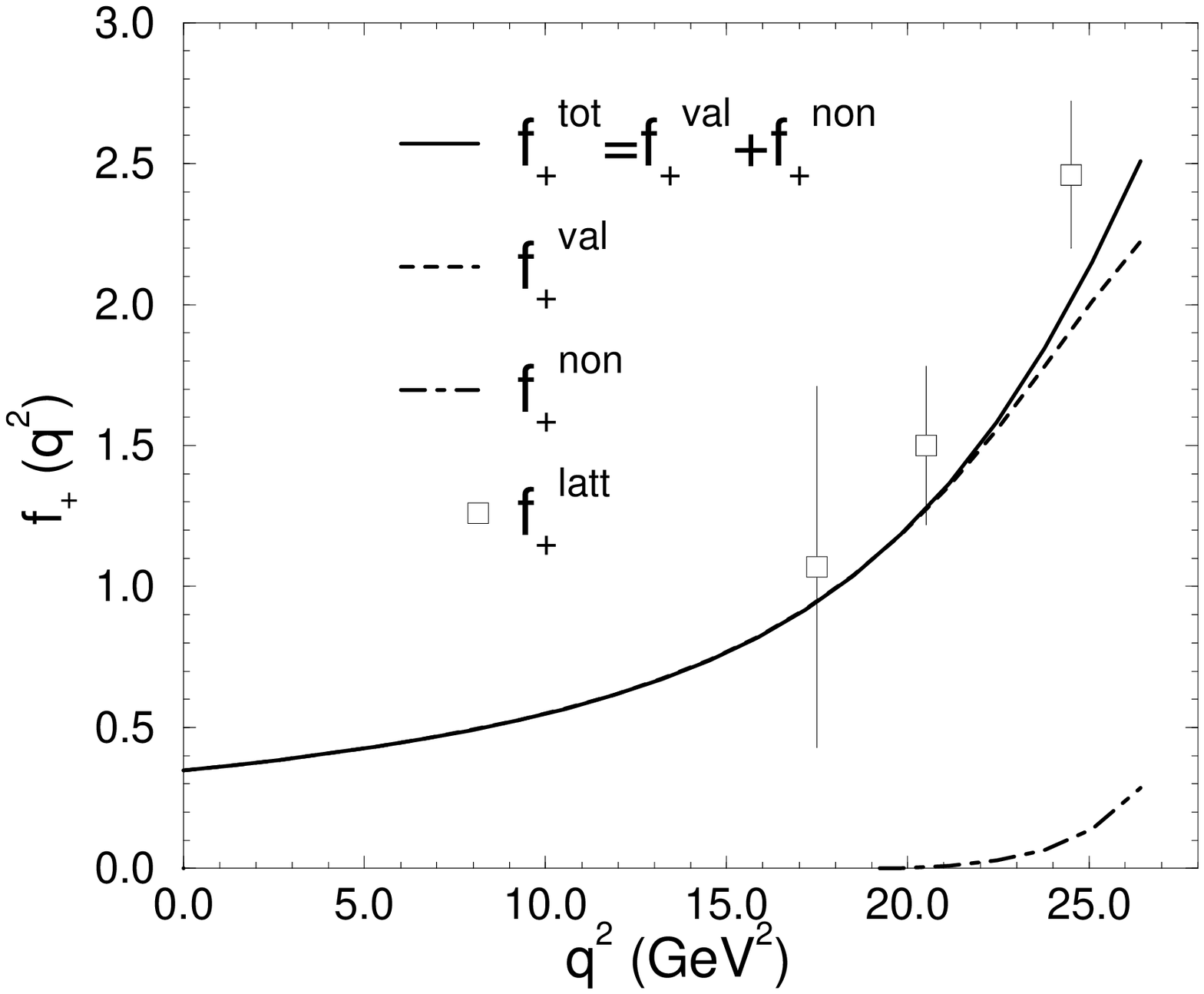}
\vskip 23.cm
\caption{}
\end{figure}

\end{document}